**On the Connection of High-Resolution NMR Spectrum Mirror Symmetry with Spin System Properties**

Dmitry A. Cheshkov[1], Dmitry O. Sinitsyn[2]

*[1]State Scientific Research Institute of Chemistry and Technology of Organoelement Compounds, 38 Shosse Entuziastov, 105118 Moscow, Russia*

*[2]Russian Center of Neurology and Neurosciences. 80 Volokolamskoe shosse, 125367 Moscow, Russia*

*E-mail: dcheshkov@gmail.com[1], d_sinitsyn@mail.ru[2]*

Moscow, 2026

**Abstract**

A connection between the symmetry of high-field NMR spectra, including higher-order spectra, and the properties of the spin system has been established. It is shown that for a spectrum to be symmetric about the mid-resonance frequency ($\nu_0$), two conditions must be satisfied: 1) the resonance frequencies of the spins must be symmetrically positioned about $\nu_0$, and 2) there must exist at least one spin ordering with a monotonic increase (or decrease) of resonance frequencies such that the spectrum is invariant under the reflection of the *J*-coupling matrix about its anti-diagonal (one way to satisfy this condition is for the *J*-coupling matrix to be explicitly persymmetric). The results were validated by calculating theoretical spectra for 3-, 4-, 5-, and 6-spin systems.

## 1. Introduction

This study investigates the spectral properties of the high-field (or 'high-resolution' in the classical sense) spin Hamiltonian under the condition $\nu \gg J$. However, for weak magnetic fields, comparable to or below the Earth's field, the following reasoning becomes inapplicable.

Some high-resolution high-field NMR spectra exhibit symmetry, which can be reasonably categorized into two types: (1) the symmetry of a first-order multiplet about its resonance frequency, and (2) the symmetry of the entire spectrum of a spin system about its spectral center of mass (mirror-symmetric spectrum), as observed, for example, in AB (Pople et al., 1959, p.122; Hoffman et al., 1971, p.57; Gunter, 2013, p.166), $A_nB_n$ (Corio, 1966; p.254), and AA'XX' (Pople et al., 1959, p.142; Hoffman et al., 1971, p.110; Gunter, 2013, p.197) spin systems. The first type of symmetry is readily explained by the fact that the coupled nuclei exist in different spin states which contribute equally to the spectral density on either side of the resonance frequency, regardless of the nuclei spin quantum numbers.

The second type of symmetry is an intrinsic property of the entire spin system. For instance, the spectra of AB (AX), $A_nB_n$ ($A_nX_n$) and AA'BB' (AA'XX') spin systems exhibit symmetry about their mid-resonance frequency, $\nu_0=(\nu_A+\nu_B)/2$. At first glance, one might expect the spectrum of the AA'A"XX'X" spin system with $C_{3V}$ symmetry (as in 1,3,5-trifluorobenzene, see Figure 1) to also be mirror-symmetric, however, this is not observed in practice (Cheshkov and Sinitsyn, 2020).

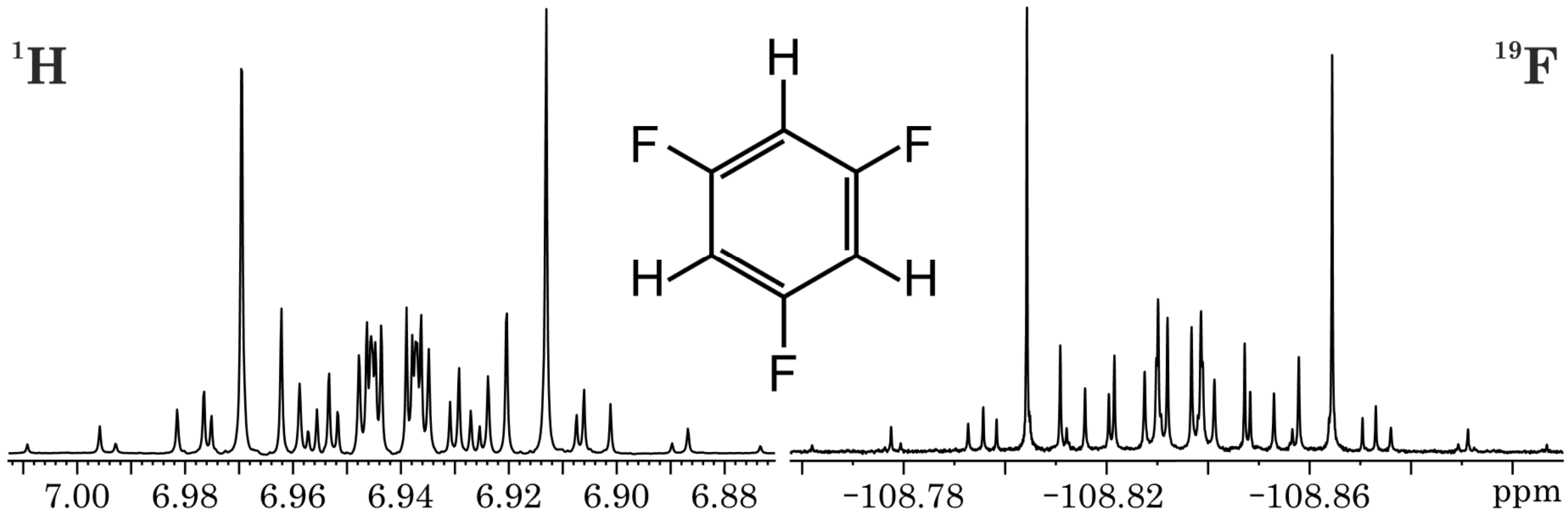

Figure 1. $^{1}$H (300.13 MHz) and $^{19}$F (282.40 MHz) spectra of 1,3,5-trifluorobenzene.

In general, high-resolution NMR spectra do not exhibit symmetry about the mid-resonance frequency. It is worth considering how this type of spectrum symmetry can arise.

## 2. Theory

Figure 2 shows the spectra of two ABC spin systems with different signs of resonance frequencies. It is readily apparent that the spectra of these spin systems are mirror images of each other. Inverting the order of the resonance frequencies while maintaining the differences between them produces a reflected spectrum, which can be thought of as reversing the direction of the frequency axis.

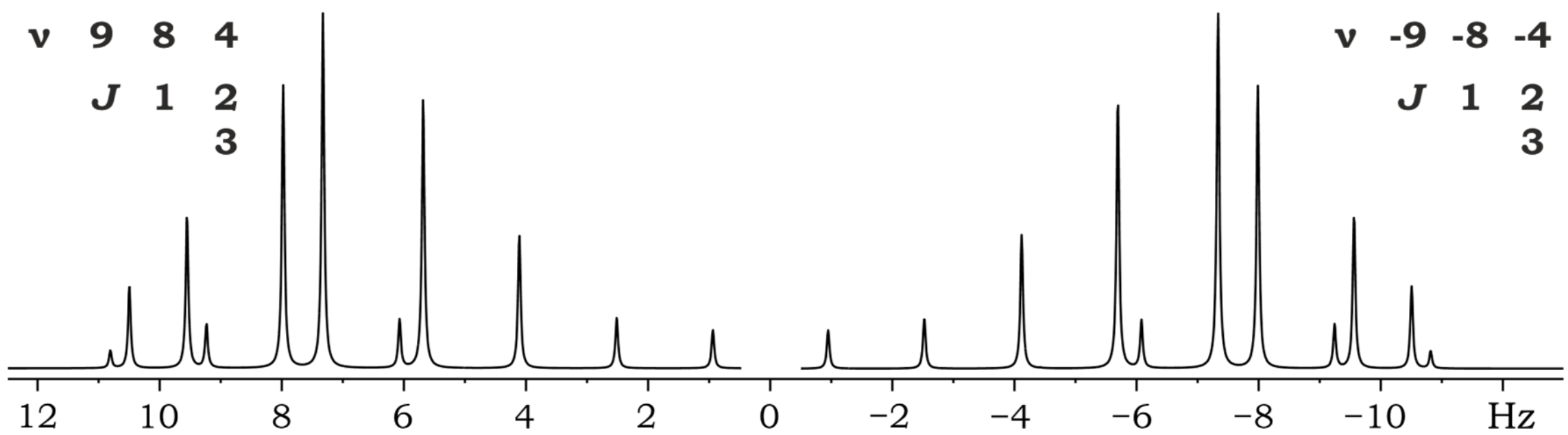

Figure 2. Spectra of two ABC spin systems, with different signs of resonance frequencies.

To demonstrate how this spectral symmetry can be achieved for a single spin system, the parameter matrices of the considered ABC spin systems are rewritten in the order of their resonance frequencies:

$$
\begin{array}{ccc}
\nu_1 & \nu_2 & \nu_3 \\
9 & 8 & 4 \\
 & J_{12} & J_{13} \\
 & 1 & 2 \\
 & & J_{23} \\
 & & 3
\end{array}
\qquad\qquad
\begin{array}{ccc}
-\nu_3 & -\nu_2 & -\nu_1 \\
-4 & -8 & -9 \\
 & J_{23} & J_{13} \\
 & 3 & 2 \\
 & & J_{12} \\
 & & 1
\end{array}
$$

It can be shown that reversing the sequence of resonance frequencies results in a reflection of the $J$-coupling matrix about its anti-diagonal, while the coupling constants on the anti-diagonal itself remain unchanged. If for a certain spin system, the parameter matrix is constructed in such a way that the resonance frequencies are symmetric about the mid-resonance frequency ($\nu_0$) and the $J$-coupling matrix is symmetric about the anti-diagonal, i.e. persymmetric (and thus bisymmetric), then reversing the order of the resonance frequencies will not alter the spectrum (Figure 3). Moreover, the spectrum itself will be symmetric about $\nu_0$. Thus, a persymmetric $J$-coupling matrix remains invariant under a reversal of the spin order. The anti-diagonal itself contains the coupling constants between spin pairs that are interchanged upon order reversal, i.e., the first and the last, the second and the penultimate, and so forth. The resonance frequencies of these spin pairs must be symmetrically arranged about the mid-resonance frequency $\nu_0$, in other words, they must be "equilibrated".

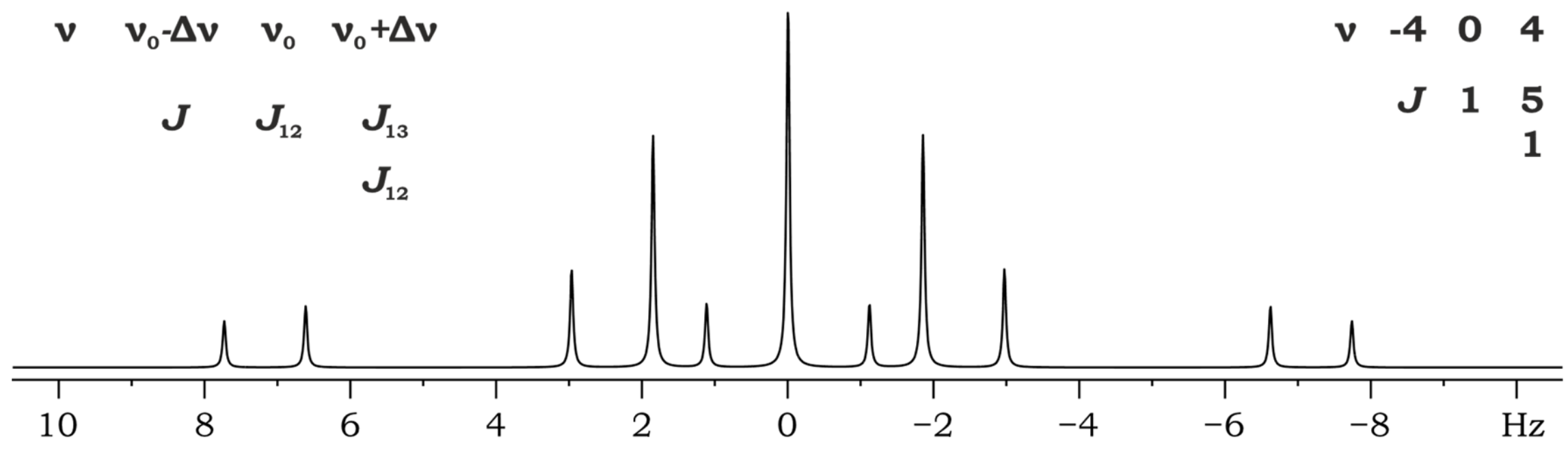

Figure 3. Mirror-symmetry spectrum of ABC spin system.

## 3. Results and discussion

Below, in Figure 4, we present several theoretically calculated mirror-symmetric spectra for spin systems with different numbers of spins (4, 5, and 6). All these spin systems possess parameter values that meet the conditions outlined above.

### 3.1. Spectral Mirror Symmetry of AA'XX' (AA'BB') and $A_nX_n$ ($A_nB_n$) Spin Systems

It is well-established that o-dichlorobenzene (ODCB) exhibits a mirror-symmetric $^1$H NMR spectrum, corresponding to an AA'XX' spin system (Figure 5).

From the algebraic properties of the AA'XX' spin system Hamiltonian, it follows that the spectrum is independent of the signs of the differences within the coupling constant pairs $\{J_{AA'}, J_{XX'}\}$ and $\{J_{AX}, J_{AX'}\}$. Consequently, the spectrum is invariant under permutation of the values within these pairs. Crucially, interchanging $J_{AA'}$ and $J_{XX'}$ is equivalent to interchanging the resonance frequencies $\nu_A$ and $\nu_X$. Since the spectrum is determined by two resonance frequencies and remains invariant under their permutation (i.e., the reversal of the resonance-frequency order), it must possess mirror symmetry.

The existence of three independent permutations within parameter pairs gives rise to eight distinct combinations of spin-system parameters corresponding to the same NMR spectrum (Fig. 5). Four of these combinations can be represented as a result of spin reordering (Fig. 5, a, c, f, h), while the others contain "non-physical" permutations of parameter values. Notably, none of these $J$-coupling matrices possess symmetry with respect to the anti-diagonal. However, $J$-coupling matrix e（Fig.5) is the anti-diagonal reflection of matrix d（Fig.5). This observation leads to more general criteria for mirror symmetry:

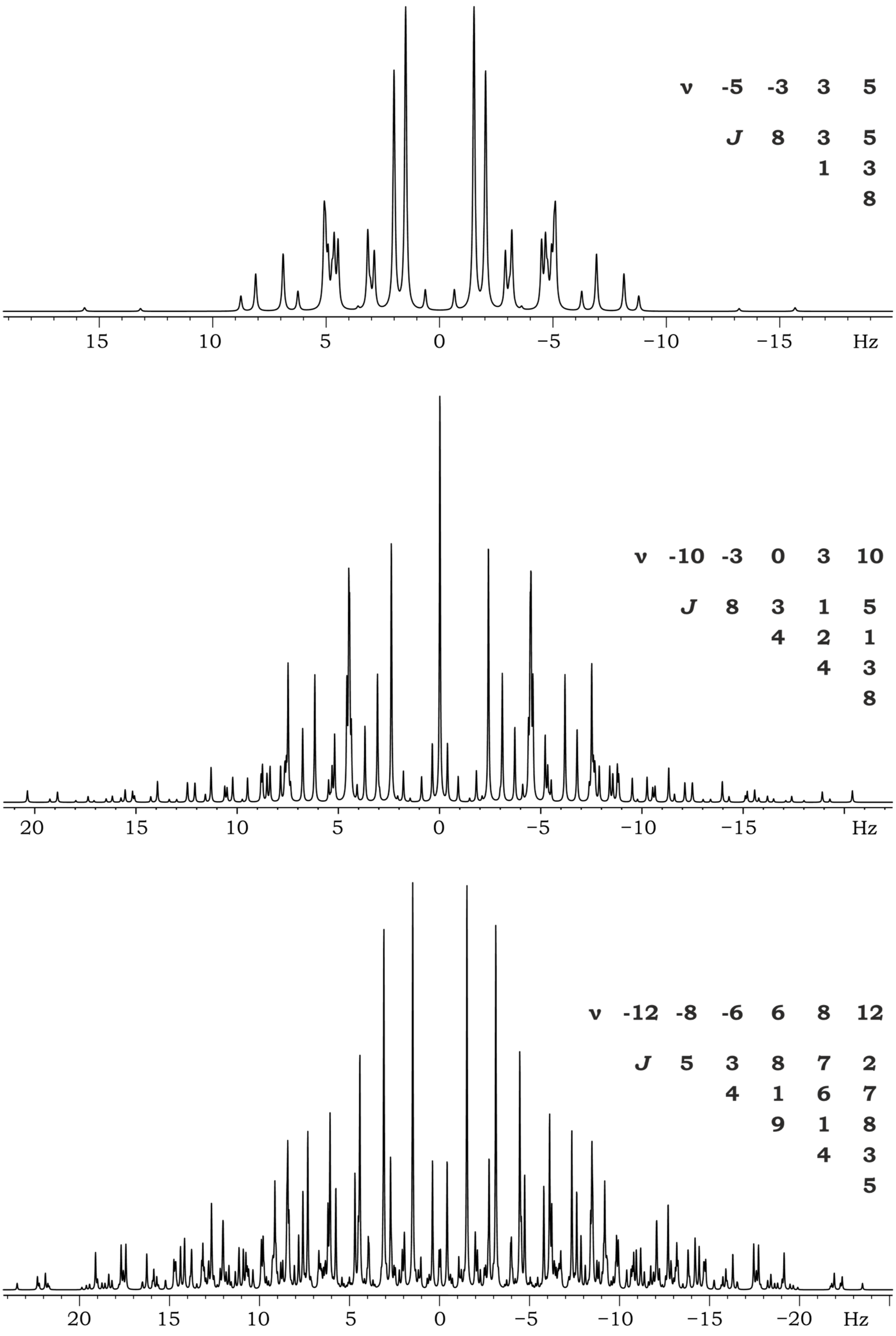

Figure 4. Some examples of theoretically calculated mirror-symmetric spectra for 4, 5 and 6-spin systems.

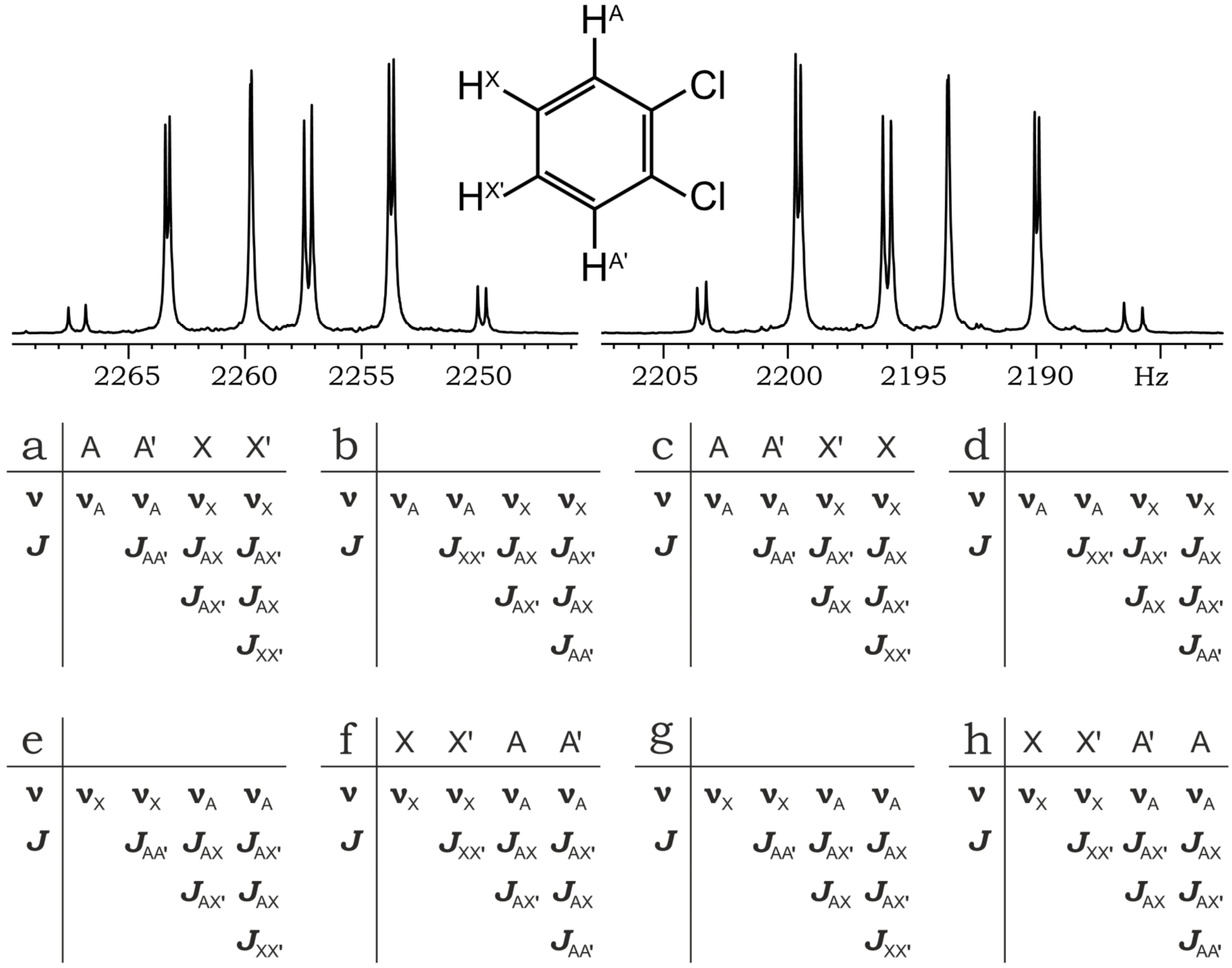

Figure 5. ODCB chemical structure with spin labeling,
$^{1}$H NMR spectrum (300.13 MHz), and spin system matrices.

If the resonance frequencies are symmetric about the mid-resonance frequency ($\nu_0$) and if there exists a spin ordering in which resonance frequencies are monotonically ordered increasingly (or decreasingly) and in this order the spectrum is invariant under the reflection of the *J*-coupling matrix about its anti-diagonal, then the spectrum will be mirror-symmetric. Alternatively: the spectrum is mirror-symmetric if it is invariant under the reflection of all resonance frequencies about their mid-resonance frequency.

The coupling constants that map onto each other above and below the anti-diagonal can either be equal or form "balanced pairs" (actually, balance each other). In fact, in the AA'XX' (AA'BB') spin system, the constants $J_{AA'}$ and $J_{XX'}$ should be considered as “equivalent” under anti-diagonal reflection. Therefore, when analyzing symmetric spin systems with chemically equivalent but magnetically non-equivalent spin groups, one must first identify all such balanced pairs of coupling constants and treat them as equivalent in assessing the symmetry of the *J*-coupling matrix.

Another family of mirror-symmetric spectra originates from spin systems with two groups of magnetically equivalent nuclei $A_nX_n$ ($A_nB_n$), the mirror symmetry of which was proved by Corio (Corio, 1966, p.254). In general, the *J*-coupling matrix in such systems lacks persymmetry because the coupling constant between the A nuclei can differ from the coupling

$D_3$-symmetric $[A'X']_3$ spin system of 1,3,5-trifluorobenzene

| | A | A' | A'' | X | X' | X'' |
|---|---|---|---|---|---|---|
| $\nu$ | $\nu_A$ | $\nu_A$ | $\nu_A$ | $\nu_X$ | $\nu_X$ | $\nu_X$ |
| $J$ | | $J_{AA'}$ | $J_{AA'}$ | $J_{AX}$ | $J_{AX'}$ | $J_{AX}$ |
| | | | $J_{AA'}$ | $J_{AX}$ | $J_{AX}$ | $J_{AX'}$ |
| | | | | $J_{AX'}$ | $J_{AX}$ | $J_{AX}$ |
| | | | | | $J_{XX'}$ | $J_{XX'}$ |
| | | | | | | $J_{XX'}$ |

$C_{2V}$-symmetric $[A'X']_4$ spin systems

| | A | A' | A'' | A''' | X | X' | X'' | X''' |
|---|---|---|---|---|---|---|---|---|
| $\nu$ | $\nu_A$ | $\nu_A$ | $\nu_A$ | $\nu_A$ | $\nu_X$ | $\nu_X$ | $\nu_X$ | $\nu_X$ |
| $J$ | | $J_{AA'}$ | $J_{AA''}$ | $J_{AA'''}$ | $J_{AX}$ | $J_{AX'}$ | $J_{AX''}$ | $J_{AX'''}$ |
| | | | $J_{AA'''}$ | $J_{AA''}$ | $J_{AX'}$ | $J_{AX}$ | $J_{AX'''}$ | $J_{AX''}$ |
| | | | | $J_{AA'}$ | $J_{AX''}$ | $J_{AX'''}$ | $J_{AX}$ | $J_{AX'}$ |
| | | | | | $J_{AX'''}$ | $J_{AX''}$ | $J_{AX'}$ | $J_{AX}$ |
| | | | | | | $J_{XX'}$ | $J_{XX''}$ | $J_{XX'''}$ |
| | | | | | | | $J_{XX'''}$ | $J_{XX''}$ |
| | | | | | | | | $J_{XX'}$ |

$D_4$-symmetric $[A'X']_4$ spin system

| | A | A' | A'' | A''' | X | X' | X'' | X''' |
|---|---|---|---|---|---|---|---|---|
| $\nu$ | $\nu_A$ | $\nu_A$ | $\nu_A$ | $\nu_A$ | $\nu_X$ | $\nu_X$ | $\nu_X$ | $\nu_X$ |
| $J$ | | $J_{AA'}$ | $J_{AA''}$ | $J_{AA'}$ | $J_{AX}$ | $J_{AX'}$ | $J_{AX'}$ | $J_{AX}$ |
| | | | $J_{AA'}$ | $J_{AA''}$ | $J_{AX}$ | $J_{AX}$ | $J_{AX'}$ | $J_{AX'}$ |
| | | | | $J_{AA'}$ | $J_{AX'}$ | $J_{AX}$ | $J_{AX}$ | $J_{AX'}$ |
| | | | | | $J_{AX'}$ | $J_{AX'}$ | $J_{AX}$ | $J_{AX}$ |
| | | | | | | $J_{XX'}$ | $J_{XX''}$ | $J_{XX'}$ |
| | | | | | | | $J_{XX'}$ | $J_{XX''}$ |
| | | | | | | | | $J_{XX'}$ |

Figure 6. Some examples of symmetric $[A'X']_n$ ($[A'B']_n$) spin systems.

constant between the X (or B) nuclei. However, the theoretical spectrum is independent of coupling between magnetically equivalent nuclei. The spectrum depends only on the inter-group couplings, which are all equal to the same value $J_{AX}$ ($J_{AB}$). Thus, the part of the *J*-coupling matrix that contains these coupling constants and defines the spectrum does indeed possess persymmetry.

### 3.2. Spectral Asymmetry of Some Symmetric $[A'X']_n$ ($[A'B']_n$) Spin Systems

Five representative examples of highly symmetric $[A'X']_n$ spin systems are shown in Figure 6. The condition of resonance frequency balance requires ordering spins by their resonance frequencies in ascending or descending order. However, within groups of chemically equivalent spins an additional ordering is required. The chemical equivalence of spin groups, determined by spin permutation symmetry, results in the topological equivalence of the coupling networks of the $[A']_n$ and $[X']_n$ subsystems. Therefore, the most symmetric form of the *J*-coupling matrix can be achieved by a consistent ordering of chemically equivalent spins from different groups such that the equality $J_{AX} = J_{A'X'} = J_{A''X''} = J_{A'''X'''}$ holds. For 1,3,5-trifluorobenzene and the 1,3,5,7-tetrafluorosubstituted cyclooctatetraene dianion, this ordering coincides with the molecular canonical topological order. Accordingly, in all examples of *J*-coupling matrices presented in Fig. 6, the chemically equivalent spins were ordered using the described scheme.

In all considered spin systems, the *J*-coupling matrices exhibit high symmetry, with the inter-group coupling blocks being persymmetric. Unlike the AA'BB' case, due to the algebraic properties of the Hamiltonian, there is no balancing of topologically equivalent homonuclear coupling constant pairs in these systems. Specifically, the constant pairs $\{J_{AA'}, J_{XX'}\}$, $\{J_{AA''}, J_{XX''}\}$, and $\{J_{AA'''}, J_{XX'''}\}$ do not form balanced pairs. Consequently, the spectra are not invariant under the permutation of constant values within these pairs nor under the permutation of $\nu_A$ and $\nu_X$, and therefore do not possess mirror symmetry, despite the high symmetry of the spin systems themselves.

Nevertheless, in the case of 1,3,5-trifluorobenzene, the high symmetry of the $[A'X']_3$ spin system results in symmetric signal patterns for the $^{1}$H(A) signal about $\nu_A$ and for the $^{19}$F(X) signal about $\nu_X$ separately.

## 4. Conclusion

The properties that a spin system must possess for its high-field NMR spectrum to be symmetric about the mid-resonance frequency $\nu_0$ have been identified. We believe these findings are of fundamental importance to the theory of spin system spectra.

All theoretical spectra were simulated using the ANATOLIA NMR software (Cheshkov et al., 2018). For an in-depth theoretical analysis of spectral mirror symmetry, see DOI: 10.48550/arXiv.2602.03871.

## 8. References

Cheshkov, D. A. and Sinitsyn, D. O., Total lineshape analysis of high-resolution NMR spectra, in: Annual Reports on NMR Spectroscopy, vol. 100, edited by: Webb, G. A., Academic Press, 61-96, https://doi.org/10.1016/bs.arnmr.2019.11.001, 2020.

Cheshkov, D. A., Sheberstov, K. F., Sinitsyn, D. O., and Chertkov, V. A.: ANATOLIA: NMR software for spectral analysis of total lineshape, Magn. Reson. Chem., 56, 449-457, https://doi.org/10.1002/mrc.4689, 2018. https://github.com/dcheshkov/ANATOLIA/

Gunter H.: NMR Spectroscopy: basic principles, concepts and applications in chemistry, 3rd ed., Wiley-VCH, ISBN: 978-3-527-33000-3, 2013.

Hoffman, R. A., Forsen, S., Gestblom, B., Analysis of NMR spectra – A guide for chemists in NMR Basic Principles and Progress, vol. 5, edited by: Diehl, P., Fluck, E., Kosfeld, R., Springer, Berlin, Heidelberg, N.Y., https://doi.org/10.1007/978-3-642-65205-9, 1971.

Corio, P. L.: Structure of High-Resolution NMR spectra, Academic Press, ISBN: 978-0124142831, 1966.

Pople, J. A., Schneider, W. G., Bernstein, H. J., High Resolution Nuclear Magnetic Resonance, McGraw-Hill, ISBN: 978-0070505162, 1959.